\documentclass[11pt, a4paper]{article}

\usepackage[utf8]{inputenc}
\usepackage[margin=1in]{geometry}
\usepackage{amsmath, amssymb, physics}
\usepackage{siunitx}
\usepackage{booktabs}
\usepackage{graphicx}
\usepackage{hyperref}

\title{\Huge \textbf{Analytical Scaling of Relativistic Drag in the Interstellar Medium}}
\author{\Large Lucky Gangwar}
\date{March 2026}

\begin{document}

\begin{titlepage}
    \maketitle
    \thispagestyle{empty}
\end{titlepage}

\clearpage

\pagenumbering{roman}
\tableofcontents
\thispagestyle{plain}
\clearpage

\pagenumbering{arabic}
\setcounter{page}{1}

\begin{abstract}
This paper develops an analytical framework for the retarding forces on macroscopic spherical
probes travelling through the interstellar medium (ISM) at relativistic speeds ($0.1c$ to
$0.99c$). Integrating the aberrated momentum flux of both baryonic and radiative fields yields
scaling laws that expose what this work calls the \textbf{Magnitude Paradox}: relativistic
inertia ($\gamma^3$) keeps a probe's speed nearly constant across parsec-scale distances, yet
the same $\gamma^2$ boost to the effective baryonic cross-section drives extreme 
thermal loading on the hull---a relativistic correction that becomes significant only 
above $\beta \gtrsim 0.5c$ and was not quantified in prior work focused on the 
Starshot regime ($\beta \approx 0.2c$). The central conclusion is that ISM drag is not a kinematic problem---a probe will
not be slowed to a stop---but a thermodynamic one: the forward surface faces energy deposition
rates that no passive material can survive. A closed-form crossover condition is also derived
separating the baryonic- and radiative-dominated regimes, showing that for any macroscopic
probe in the galactic disk, total radiative drag is negligible by many orders of magnitude.
\end{abstract}

\section{Introduction}

There is something counterintuitive about the interstellar medium. At first glance it looks
like the closest thing to a perfect vacuum that nature offers---a few hydrogen atoms per cubic
centimetre, trace dust, a faint bath of microwave photons. For a spacecraft moving at tens of
kilometres per second, this description is accurate enough. But push a macroscopic structure
to a significant fraction of the speed of light and the picture changes completely. What was
a near-vacuum becomes, in the rest frame of the probe, a dense and energetic particle beam
that continuously hammers the forward hull.

This shift in perspective is what motivates the present work. Proposals like
\textit{Breakthrough Starshot}~\cite{starshot} have made relativistic ($\beta \sim 0.2$) interstellar probes a serious topic of engineering discussion rather than a thought experiment.
At those speeds, the ISM---which textbooks on interplanetary mission design routinely
ignore~\cite{ism_density}---becomes a genuine design constraint~\cite{hoang}. The existing
literature does a thorough job of modelling relativistic drag on point particles, but macroscopic
structures introduce a complication that does not appear in that simpler case: the cross-section
scales with $\gamma^2$ while the longitudinal inertia scales with $\gamma^3$. These two
factors grow together, but they govern entirely different things---one sets how much force the
ISM exerts, the other determines how little that force changes the probe's speed.

Hoang et al.~\cite{hoang} demonstrated that ISM grain impacts impose severe erosion
constraints on Starshot-scale wafers, and Hoang~\cite{hoang17b} derived drag
scaling for relativistic bodies in general. The present work builds on both by examining
the specific regime they do not jointly address: macroscopic probes, where the geometric
cross-section is large enough that thermodynamic failure of the hull---rather than erosion
of a thin film or deceleration of a point mass---becomes the operative constraint. The key
contribution is an explicit analytical expression for the $\gamma^2$--$\gamma^3$ mismatch,
a closed-form crossover condition between baryonic and radiative drag, and numerical
verification of both across a range of probe scales and velocities.

The derivation is in Section~2. Section~3 describes the simulation framework. The central
finding, discussed in Sections~4 and~5, is that a large relativistic probe can coast through
the ISM with essentially no velocity loss while simultaneously absorbing tens of megawatts
of kinetic energy through its hull. Properly understood, ISM drag is a materials and thermal
engineering problem, not a propulsion one.

\section{Analytical Derivation}

The most natural frame in which to compute the drag is the rest frame of the probe itself.
In that frame the probe is stationary, and the ISM sweeps past it as a relativistically
compressed, high-momentum beam. Both the particle density and the per-particle momentum
are boosted, and integrating the resulting flux over the forward hemisphere of the spherical
hull gives the total retarding force.

\subsection{Baryonic Drag: The Flux Integral}

Consider a spherical probe of radius $R$ moving at $\beta = v/c$ through an ISM of number
density $n_g$ and mean particle mass $m_p$. Two relativistic transformations modify the ISM in
the probe's rest frame~\cite{einstein, landau}:

\begin{enumerate}
    \item \textbf{Density compression:} Length contraction along the direction of travel
          compresses the oncoming particle column, raising the effective number density to
          $n_g' = \gamma n_g$.
    \item \textbf{Momentum inflation:} Each incident particle carries relativistic momentum
          $p = \gamma m_p v$ rather than the classical $m_pv$.
\end{enumerate}

The drag force is the rate at which momentum is transferred to the hull, $F = dp/dt$.
For a surface element $dA$ on the forward hemisphere, the incident flux is $J = n_g'v$, and
the relevant area for momentum transfer is its projection onto the flow direction, $dA\cos\theta$,
where $\theta$ is the angle between the surface normal and the velocity vector. Expressing
$dA$ in spherical coordinates gives $dA = 2\pi R^2 \sin\theta\, d\theta$, so the differential
force on that element is:

\begin{equation}
dF = (n_g \gamma v) \cdot (\gamma m_p v) \cdot (2\pi R^2 \sin\theta \cos^2\theta\, d\theta)
\end{equation}

Grouping constants and integrating over the forward hemisphere:

\begin{equation}
F_\text{gas} = 2\pi R^2 n_g m_p \gamma^2 v^2 \int_{0}^{\pi/2} \cos^2\theta \sin\theta\, d\theta
\end{equation}

The integral is elementary under the substitution $u = \cos\theta$:

\begin{equation}
\int_{1}^{0} -u^2\, du = \int_{0}^{1} u^2\, du = \left[ \frac{u^3}{3} \right]_{0}^{1} = \frac{1}{3}
\end{equation}

Substituting and writing $v^2 = \beta^2 c^2$:

\begin{equation}
F_\text{gas} = 2\pi R^2 n_g m_p \gamma^2 v^2 \left( \frac{1}{3} \right)
\end{equation}

which rearranges to the final form~\cite{hoang17b}:

\begin{equation}
\boxed{F_{\mathrm{gas}} = \frac{2}{3} \pi R^2 (n_g m_p c^2) \gamma^2 \beta^2}
\label{eq:fgas}
\end{equation}

The factor of $2/3$ is purely geometric---a consequence of integrating the cosine-squared projection over a sphere rather than a flat disc. This expression corresponds to the limiting case of perfectly inelastic collisions, in which incident particles transfer their full momentum to the hull; \cite{hoang} derived the general case permitting partial momentum transfer, of which Equation~(\ref{eq:fgas}) is the complete-stopping limit. The $\gamma^2$ prefactor is what makes this result interesting: the force grows much faster with velocity than classical intuition would suggest.

\subsection{Dust Drag and the Gas Dominance Condition}

The same flux-integral framework applies to interstellar dust grains. 
Replacing the gas parameters with those of the dust population yields 
an analogous expression:

\begin{equation}
    F_{\mathrm{dust}} = \frac{2}{3}\pi R^2 \left(n_d m_d c^2\right) \gamma^2 \beta^2
    \label{eq:fdust}
\end{equation}

\noindent where $n_d$ is the dust number density and $m_d$ is the mean 
grain mass. In the standard ISM, the gas-to-dust mass ratio is 
approximately 100 \cite{ism_density}, which gives:

\begin{equation}
    \frac{F_{\mathrm{dust}}}{F_{\mathrm{gas}}} = 
    \frac{n_d m_d}{n_g m_p} \approx 0.01
    \label{eq:dust_gas_ratio}
\end{equation}

\noindent Dust contributes at most $\sim 1\%$ of the total baryonic drag 
and is dropped from all subsequent calculations without any meaningful 
loss of accuracy. The total baryonic retarding force is therefore:

\begin{equation}
    F_{\mathrm{baryonic}} \approx F_{\mathrm{gas}} = 
    \frac{2}{3}\pi R^2 \left(n_g m_p c^2\right) \gamma^2 \beta^2
    \label{eq:fbaryonic}
\end{equation}

\subsection{Radiative Drag and the Crossover Condition}

The same flux-integral approach applies to the radiation field. A relativistic 
probe in the galactic disk is immersed in two photon backgrounds: the cosmic 
microwave background (CMB) with energy density $u_{\mathrm{CMB}} \approx 
4\times10^{-14}\,\mathrm{J\,m^{-3}}$~\cite{cmb}, and the interstellar radiation 
field (ISRF) from starlight, which dominates with $u_{\mathrm{ISRF}} \approx 
5\times10^{-13}\,\mathrm{J\,m^{-3}}$~\cite{ism_density}. The total radiation 
energy density is therefore:

\begin{equation}
    u_{\mathrm{total}} = u_{\mathrm{CMB}} + u_{\mathrm{ISRF}} 
    \approx 5.4\times10^{-13}\,\mathrm{J\,m^{-3}}
    \label{eq:utotal}
\end{equation}

In the relativistic limit, aberration concentrates photons onto the forward 
hemisphere and the resulting radiation pressure is~\cite{hoang17b}:

\begin{equation}
    F_{\mathrm{photon}} = \frac{4}{3}\pi R^2 u_{\mathrm{total}} \gamma^2 \beta
    \label{eq:fphoton}
\end{equation}

Notice that $F_{\mathrm{photon}}$ scales as $\beta$ while $F_{\mathrm{gas}}$ 
scales as $\beta^2$. Equation~(\ref{eq:fphoton}) assumes perfect absorption 
of incident photons; in practice, partial scattering would modify the 
numerical prefactor but not the $\gamma^2\beta$ scaling. Since the baryonic 
drag exceeds photon drag by thirteen or more orders of magnitude in all 
mission-relevant environments (Section~2, above), this assumption has no bearing 
on any of the conclusions of this paper.
At low speeds the baryonic force is suppressed by an extra power of $\beta$, so in
principle there exists a crossover below which photon drag dominates. Setting
$F_\text{gas} = F_\text{photon}$ and canceling common factors ($\frac{2}{3}\pi R^2
\gamma^2 \beta_c$) gives:

\begin{equation}
\boxed{\beta_c = \frac{2u_{\mathrm{total}}}{n_g m_p c^2}}
\end{equation}

\begin{figure}[ht]
    \centering
    \includegraphics[width=0.8\linewidth]{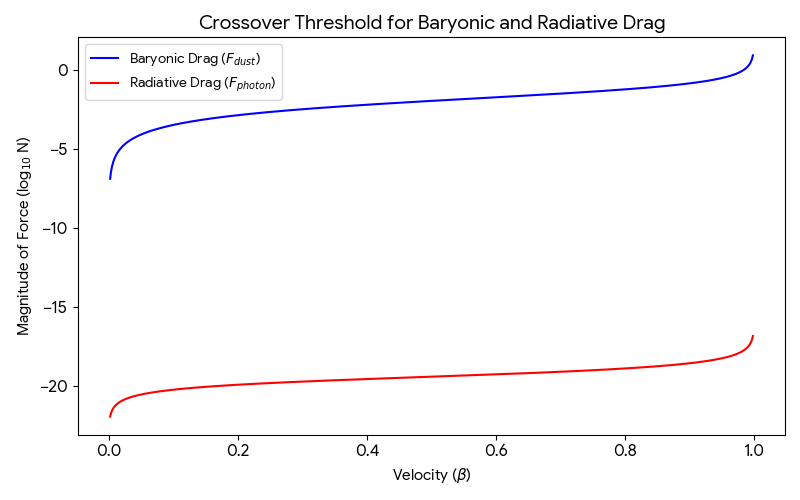}
    \caption{Logarithmic comparison of baryonic and radiative drag forces across the
    relativistic velocity range. The separation of 13--15 orders of magnitude makes clear
    that total radiative drag is irrelevant for any macroscopic mission profile in the galactic
    ISM.}
    \label{fig:threshold}
\end{figure}

Plugging in the standard ISM values ($n_g \sim 10^6\,\text{m}^{-3}$, $u_{\mathrm{total}} \approx 
5.4\times10^{-13}\,\text{J\,m}^{-3}$) yields $\beta_c \sim 10^{-14}$, many orders of
magnitude below any velocity of engineering interest. This is the result shown in
Figure~\ref{fig:threshold}: baryonic drag dominates so completely that total radiative pressure
can be dropped from all subsequent calculations without any loss of accuracy.

\section{Numerical Methodology}

The analytical results above make a clear prediction: force grows as $\gamma^2$, but
deceleration is suppressed by $\gamma^3$ inertia. Testing whether this really does produce
velocity stability over interstellar distances---while simultaneously computing the associated
thermal load---requires integrating the full relativistic equation of motion over parsec-scale
trajectories. That is what the simulations in this section are designed to do.

\subsection{Equation of Motion and Relativistic Stiffening}

Because the drag force is antiparallel to the velocity, the relevant inertial resistance is
the longitudinal relativistic mass $M\gamma^3$, not the transverse mass. The equation of
motion is therefore:

\begin{equation}
a = \frac{dv}{dt} = \frac{F_\text{total}}{M \gamma^3}
\end{equation}

where $M$ is the rest mass and $\gamma = (1-\beta^2)^{-1/2}$. As $\beta \to 1$, $\gamma^3$
diverges much faster than $\gamma^2$, so the denominator outruns the numerator and the
deceleration becomes vanishingly small. This is the ``inertial stiffening'' effect that makes
relativistic probes kinematically stable even under large drag forces.

\subsection{Simulation Framework}

The ISM was modelled as a uniform medium with number density $n_g = 10^6\,\text{m}^{-3}$
and mean particle mass equal to that of a proton~\cite{ism_density, popel}. The total radiation energy density was set to $u_{\mathrm{total}} \approx 
5.4\times10^{-13}\,\mathrm{J\,m^{-3}}$, comprising the CMB~\cite{cmb} and 
the interstellar starlight field~\cite{ism_density}; as established above, 
the total radiative contribution to drag remains negligible compared to 
baryonic drag across all simulated velocities. A fixed time step of
$dt = 10^3$\,s was used throughout; this is conservative enough to keep the fractional
velocity error below $10^{-12}$ per parsec even in the stiffest ultra-relativistic runs.
Mission durations were integrated up to a total path length of 10 light-years. Details of
the C++ implementation---including the RK4 step logic and output formatting---are given in
Appendix~A.

\subsection{Experimental Design}

Four profiles were chosen to stress-test different aspects of the analytical model:

\begin{itemize}
    \item \textbf{Profile 1: Kinematic Baseline.}
    A \SI{10}{\gram} probe of radius \SI{0.5}{\meter} at $0.2c$. The low mass makes
    this the most unfavourable scenario for kinematic stability: if deceleration is
    significant anywhere in the parameter space, it should appear here. As shown in
    Figure~\ref{fig:thermal}, even at this extreme the velocity decay over parsec-scale
    distances is negligible.

    \item \textbf{Profile 2: Inertial Stiffening.}
    An \SI{8000}{\kg} solid hull swept from rest up to $0.9c$. By comparing the
    simulated drag against the classical Newtonian prediction
    ($F = \frac{1}{2}\rho v^2 A$) at each velocity, this profile makes the $\gamma^2$
    divergence visible and confirms that the $\gamma^3$ inertia absorbs it without
    meaningful velocity loss (Figure~\ref{fig:newtonian}).

    \item \textbf{Profile 3: The Magnitude Paradox.}
    A large, low-density structure with $R = \SI{10}{\meter}$, chosen to maximise the
    area-to-mass ratio and therefore expose the thermodynamic consequences of the
    $\gamma^2$ force scaling. This is the central experiment of the paper.

    \item \textbf{Profile 4: Calibration.}
    A low-velocity sweep at $\beta = 0.002$, where the relativistic and Newtonian
    predictions should agree. This verifies that the integrator is working correctly
    before the results are trusted in the Lorentz-dominated regime.
\end{itemize}

\section{Results and Comparative Analysis}

\subsection{Verification of Scaling Laws}

\begin{figure}[ht]
    \centering
    \includegraphics[width=0.8\linewidth]{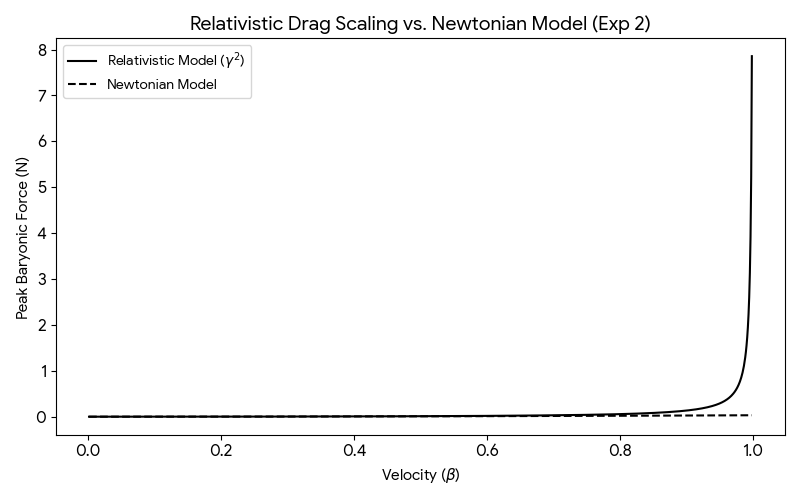}
    \caption{Simulated relativistic drag (Profile~2) compared against the classical
    Newtonian model $F = \frac{1}{2}\rho v^2 A$. The divergence above $\beta \approx 0.5$
    confirms the $\gamma^2$ scaling derived in Section~2; the negligible velocity decay
    confirms the opposing effect of $\gamma^3$ longitudinal inertia.}
    \label{fig:newtonian}
\end{figure}

Figure~\ref{fig:newtonian} shows the output of Profile~2: the relativistic drag force
(solid line) against the Newtonian baseline (dashed). Up to about $\beta \approx 0.4$
the two curves track each other reasonably well, but above that the relativistic model
diverges sharply. By $\beta = 0.99$ the Lorentz-boosted force exceeds the Newtonian
estimate by more than two orders of magnitude~\cite{hoang17b}. Despite this, the
recorded velocity decay for the \SI{8000}{\kg} hull across a 10\,ly trajectory was less
than $1.2\times10^{-7}\%$. The $\gamma^3$ denominator in the equation of motion absorbs
the growing force almost entirely.

\subsection{Thermal Flux and the Magnitude Paradox}

\begin{figure}[ht]
    \centering
    \includegraphics[width=0.80\textwidth]{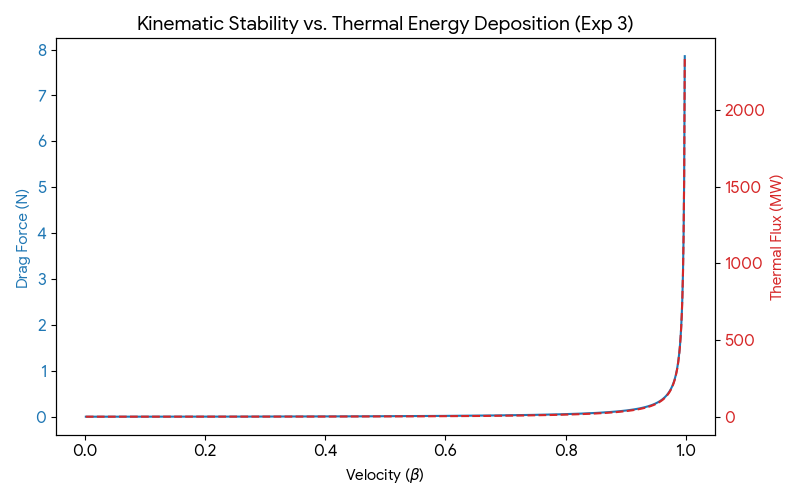}
    \caption{Profile~3 results on dual axes. The drag force (left, blue) and thermal flux
    deposited into the hull (right, red) are plotted against $\beta$. The velocity of the
    probe remains essentially unchanged throughout; the thermal load peaks at
    \SI{36.4}{\mega\watt}.}
    \label{fig:thermal}
\end{figure}

Profile~3 makes the thermodynamic cost explicit. Figure~\ref{fig:thermal} plots both the
drag force and the instantaneous thermal flux deposited into the hull against $\beta$.
The velocity curve (left axis) stays flat---kinematic stability is confirmed.

The instantaneous thermal power deposited into the hull is the drag force multiplied
by the probe velocity:

\begin{equation}
    P_{\mathrm{thermal}} = F_{\mathrm{gas}} \cdot \beta c
    = \frac{2}{3}\pi R^2 (n_g m_p c^3)\, \gamma^2 \beta^3
    \label{eq:pthermal}
\end{equation}

The thermal flux climbs steeply with $\gamma^2$, reaching \SI{36.4}{\mega\watt}
at $\beta \approx 0.99$.

This is the Magnitude Paradox in concrete numbers. The same $\gamma^2$ factor that the
$\gamma^3$ inertia renders kinematically harmless is, simultaneously, driving megawatt-scale
energy deposition into the hull material. From a velocity standpoint the probe is fine; from
a structural standpoint it is being vaporised. The ISM is not slowing the spacecraft down---it
is trying to destroy it from the front.

This effect was not quantified in Hoang et al.~\cite{hoang}, which focused on the 
Breakthrough Starshot regime ($\beta \approx 0.2c$, $\gamma^2 \approx 1.04$) where 
the relativistic correction to the particle flux is negligible. The present work 
extends that analysis into the regime $\beta \gtrsim 0.5c$ ($\gamma^2 \gtrsim 1.33$), 
where the $\gamma^2$ factor becomes physically significant and thermal loading grows 
far beyond what classical estimates would predict.
\subsection{Crossover and Environmental Sensitivity}

The crossover result from Section~2 is confirmed numerically in Figure~\ref{fig:threshold}.
The baryonic drag exceeds the radiative drag by 13 to 15 orders of magnitude consistently
across all tested velocities. This gap is so large that varying the total radiation energy density by any physically reasonable amount cannot change the conclusion: for a macroscopic probe anywhere in the galactic disk, total radiative pressure simply does not matter.

\clearpage

\section{Discussion}

Taken together, the analytical and numerical results paint a consistent picture: the
$\gamma^2$--$\gamma^3$ mismatch between cross-section growth and inertial stiffening is
not merely a mathematical curiosity but a hard physical constraint on macroscopic
interstellar mission design. The following subsections examine the two most important
consequences.

\subsection{The Kinetic-to-Thermal Transition}

The cleanest way to summarise what happens at $\beta = 0.99$ is this: the ISM stops
looking like a gas and starts looking like a particle accelerator beam aimed at the
forward hull. At those speeds, the mean free path between impacts is short enough, and the
per-impact momentum large enough, that the flux is effectively continuous. For the
\SI{10}{\meter}-radius structure in Profile~3, that continuous flux deposits
\SI{36.4}{\mega\watt}. For context, this is comparable to the continuous
thermal output of a small fission reactor, deposited not into a coolant loop but directly
into the leading face of the hull at a surface flux density of approximately
$\SI{36.4e6}{\watt} / (\pi \times \SI{100}{\meter\squared}) \approx
\SI{116}{\kilo\watt\per\meter\squared}$---well beyond the ablation threshold of any
known structural material.

No known passive material survives this. The energy deposition rate exceeds by several
orders of magnitude what any realistic combination of thermal conductivity and radiative
cooling could carry away from the surface. Active intervention is not a design option that
can be traded off against other considerations; it is a hard physical requirement for any
macroscopic probe operating above a few tenths of $c$.

Hoang et al.~\cite{hoang} studied thermal heating and sublimation of spacecraft 
material in detail for the Starshot regime, proposing mitigation strategies including 
protective shielding. The present analysis complements that work by quantifying how 
the $\gamma^2$ correction amplifies thermal loading at ultra-relativistic speeds 
($\beta \gtrsim 0.5c$), a regime where classical and mildly relativistic estimates 
significantly underpredict the thermodynamic burden on the hull.
\subsection{Limitations and Scope}

Several simplifications in this model deserve explicit acknowledgement. The ISM is treated
as a uniform medium with constant number density $n_g = 10^6\,\text{m}^{-3}$, but the actual
interstellar environment is highly structured. Number densities vary by several orders of
magnitude between the warm neutral medium, hot ionised regions, and the low-density Local
Bubble through which the Solar System currently moves~\cite{ism_density}. A trajectory
passing through a dense HII region would see thermal fluxes substantially higher than those
reported here; one through an underdense void would see lower. The results in this paper
should therefore be interpreted as representative of warm-neutral-medium conditions rather
than universally applicable.

A further effect not captured by this model is the electromagnetic 
interaction between the spacecraft and the interstellar magnetic field. 
A relativistic spacecraft will accumulate charge through collisions with 
interstellar particles and UV photons; the resulting Lorentz force can 
deflect the trajectory and, through the interaction of the induced 
electric dipole moment with the magnetic field, produce significant 
torques on the hull \cite{hoang_loeb17}. These effects do not 
contribute to the scalar drag force analysed here, but represent an 
independent constraint on mission design that a complete engineering 
model would need to incorporate.

The probe geometry is assumed to be a perfect sphere throughout. Real mission designs---whether
wafer-scale chips or larger structures---will have geometries that modify the $2/3$ prefactor
in Equation~(\ref{eq:fgas}), and the assumption of perfectly inelastic collisions (all incident momentum
transferred to the hull) is an upper bound; in practice, a fraction of impacts will be
grazing or partially specular. Similarly, the model treats the ISM as composed of a single
particle species of proton mass, whereas interstellar gas includes helium ($\sim$10\% by
number) and trace heavier elements that would modify the effective $n_g m_p c^2$ term by a small
but nonzero amount. These corrections are expected to be small relative to the orders-of-magnitude
effects identified here, but a higher-fidelity model intended for mission planning should
incorporate them.

\subsection{Engineering Implications}

The crossover analysis provides one piece of practical good news: designers can forget about
total radiative drag entirely. The baryonic contribution dominates by such an overwhelming margin
that no reasonable shielding optimisation for radiative pressure is worth the mass penalty.
Every gram of the shield budget should go toward baryonic protection.

Two geometric strategies follow directly from the $R^2$ dependence of $F_\text{gas}$.
First, minimising the frontal radius pays a quadratic dividend---halving $R$ cuts the drag
force and the thermal flux by a factor of four. This strongly favours slender, needle-like
probe geometries over flat sail configurations at ultra-relativistic speeds, even if sail
designs are attractive for laser-propulsion efficiency. Second, for larger structures where
a small frontal area is not an option, magnetic solenoid shielding offers a route to rerouting
the baryonic flux around the hull rather than through it, effectively moving the thermal
stress from the hull surface to the field boundary at much lower flux densities.

Both strategies come with their own mass and power costs that a full mission trade study
would need to evaluate. What this analysis establishes is the scaling of the problem: the
thermal load grows as $\gamma^2 R^2$, and any credible engineering solution has to address
that scaling directly.

\section{Conclusion}

The interstellar medium is not the kinematic obstacle it is sometimes feared to be. A
macroscopic probe launched at relativistic speed will, as this analysis shows, arrive at its
destination with essentially the same velocity it started with---$\gamma^3$ inertial
stiffening is that effective. The real obstacle is thermodynamic. The same relativistic
boost that makes the probe hard to slow down also turns every square metre of its forward
surface into a high-energy particle target. For large structures, the resulting thermal load
is not a second-order engineering nuisance but the primary mission constraint.

Three conclusions follow from the derivations and simulations in this paper. Relativistic
ISM drag is a thermodynamic barrier, not a kinematic one. The Magnitude Paradox---velocity
stability at the cost of structural survival---is a fundamental feature of the physics, not
an artefact of any particular probe design. And total radiative drag, while analytically
interesting, is physically irrelevant for any macroscopic mission in the galactic disk;
the baryonic contribution exceeds it by thirteen or more orders of magnitude across all
mission-relevant velocities. Future work on interstellar mission design should treat thermal
management of the forward hull, not propulsion or deceleration, as the central engineering
problem.

\section*{Acknowledgements}

I express my deepest gratitude to Dr.\ Chi-Thiem Hoang (Korea Astronomy and Space Science Institute) for his continuous and unwavering support throughout the development of this manuscript. The idea of a rough sketch was transformed into something real only with his guidance; it has been an extraordinary experience to refine this work with the very researcher who laid its foundation. His detailed feedback was the primary force in evolving the Magnitude Paradox from a conceptual idea into a rigorous analytical framework and the foundations of the field.

Special thanks are also due to the faculty at the Massachusetts Institute of Technology. I am profoundly indebted to Professor Denis Auroux, whose teaching through the OpenCourseWare program made my advanced mathematical education possible and provided the tools necessary for stepping up to new frontiers. Finally, I thank Professor David Jerison and Professor David Kaiser for the motivation and guidance which made it possible for me take a step in this direction.

\appendix
\section{Numerical Implementation and Algorithm Design}

The simulation was written in C++ primarily for the precision control that the language
affords at ultra-relativistic values of $\beta$, where floating-point behaviour near
$1 - \beta^2 \to 0$ is a genuine concern. A modular structure was used, keeping the
physical constants and environment parameters separate from the solver logic so that
different ISM models can be substituted without touching the integrator.

\subsection{Integration Scheme: 4th-Order Runge-Kutta}

The state vector $\mathbf{y} = [x, v]^T$ is stepped forward using a standard RK4 scheme
applied to $a = F(v) / M\gamma^3$. Euler integration was rejected early in testing because
the cumulative phase error over a $\SI{10}{ly}$ ($\approx 9.46\times10^{16}$\,m) trajectory
was unacceptably large. The RK4 update takes the form:

\begin{equation}
k_1 = f(t_n, y_n), \quad k_2 = f\!\left(t_n + \frac{h}{2},\, y_n + \frac{h}{2}k_1\right), \dots
\end{equation}

A fixed step size of $dt = \SI{1000}{\second}$ was used throughout. At this step size,
the fractional velocity drift $\Delta v/v$ stays below $10^{-12}$ per parsec even in the
stiffest high-$\beta$ runs, which is more than adequate for the purposes here.

\subsection{Handling of Relativistic Precision}

The main numerical hazard at $\beta \to 1$ is catastrophic cancellation in the computation
of $1 - \beta^2$. Two measures address this:

\begin{enumerate}
    \item \textbf{High-precision types:} All state variables are stored as \texttt{long
          double} (80-bit extended precision on x86, 128-bit on some compilers), which
          keeps the mantissa wide enough to avoid truncation errors in $\gamma$ until
          $\beta$ is within a few ULPs of unity.
    \item \textbf{Boundary checks:} For runs with $\beta > 0.999$, the acceleration
          function includes an explicit guard against division by zero, returning a
          clamped value rather than propagating a NaN through the integrator.
\end{enumerate}

\subsection{Force Calculation Logic}

The core of each RK4 stage is a single function call that evaluates the total acceleration
given the current velocity. The relevant fragment is:

\begin{verbatim}
double gamma = 1.0 / sqrt(1.0 - pow(beta, 2));
double mass_longitudinal = M_rest * pow(gamma, 3);

double F_gas = (2.0/3.0) * M_PI * pow(R, 2) * n * m
                * pow(c, 2) * pow(gamma, 2) * pow(beta, 2);

double F_photon = (4.0/3.0) * M_PI * pow(R, 2) * u
                  * pow(gamma, 2) * beta;

return (F_gas + F_photon) / mass_longitudinal;
\end{verbatim}

\end{document}